\documentstyle [12pt,epsf] {article}

\catcode`@=12
\newcommand{\beq}{\begin{equation}}
\newcommand{\eeq}{\end{equation}} 
\newcommand{\bea}{\begin{eqnarray}}
\newcommand{\eea}{\end{eqnarray}} 
\newcommand{\bers}{\begin{eqnarray*}}
\newcommand{\eers}{\end{eqnarray*}}
\newcommand{\bt}{\begin{itemize}} 
\newcommand{\et}{\end{itemize}} 

\def\sss{\scriptscriptstyle}
\def\sss{\scriptscriptstyle}
\def\bd{B_d^0}
\def\bdbar{{\bar B}^0_d}
\def\bs{B_s^0}

\def\barp{{\raise.35ex\hbox
{${\sss (}$}}---{\raise.35ex\hbox{${\sss )}$}}}
\def\bdbarp{\hbox{$B_d$\kern-1.4em\raise1.4ex\hbox{\barp}}}
\def\bsbarp{\hbox{$B_s$\kern-1.4em\raise1.4ex\hbox{\barp}}}

\def\kbar{{\bar K}^0}
\def\kstarbar{{\bar K}^*}

\def\bdbarp{\hbox{$B_d$\kern-1.4em\raise1.4ex\hbox{\barp}}}
\def\bsbarp{\hbox{$B_s$\kern-1.4em\raise1.4ex\hbox{\barp}}}

\def\kbar{{\bar K}^0}
\def\kstarbar{{\bar K}^*}

\def\roughly#1{\mathrel{\raise.3ex\hbox
{$#1$\kern-.75em\lower1ex\hbox{$\sim$}}}}

\def\bra#1{\langle#1|}
\def\ket#1{|#1\rangle}

\def\ZI{Z_{\sss I}}
\def\ZR{Z_{\sss R}}

\def\npb#1#2#3{{\it Nucl.\ Phys.} {\bf B#1}, #3 (#2)}
\def\plb#1#2#3{{\it Phys.\ Lett.} {\bf #1B}, #3 (#2)}
\def\prd#1#2#3{{\it Phys.\ Rev.} {\bf D#1}, #3 (#2)}
\def\newprd#1#2#3{{\it Phys.\ Rev.} {\bf D#1}: #3 (#2)}

\def\prl#1#2#3{{\it Phys.\ Rev.\ Lett.} {\bf #1}, #3 (#2)}

\begin{document} 
\vspace{0.5in} 
\oddsidemargin -.375in 
\newcount\sectionnumber 
\sectionnumber=0 
\def\bra#1{\left\langle #1\right|} 
\def\ket#1{\left| #1\right\rangle} 
\def\be{\begin{equation}} 
\def\ee{\end{equation}} 
\thispagestyle{empty} 
\begin{flushright}  
UdeM-GPP-TH-02-105\\
\end{flushright}
\vskip0.5truecm

\begin{center} 

{\large \bf
\centerline{
$B_{d,s}^0 \rightarrow K^{(*)} \bar{K}^{(*)}$: CP phase $\alpha$ and New
Physics {\footnote{ Talk presented by Alakabha Datta at the 5th International Conference on Hyperon, Charm and Beauty Hadrons.}}}}
\vspace*{1.0cm}
%\vskip1cm

{\large Alakabha Datta and David London} \vskip0.3cm
{\it  Laboratoire Ren\'e J.-A. L\'evesque, Universit\'e de
  Montr\'eal,} \\
{\it C.P. 6128, succ.\ centre-ville, Montr\'eal, QC, Canada H3C 3J7} \\
\vskip0.5cm
\bigskip
(\today)
\vskip0.5cm
{\Large Abstract\\}
\vskip3truemm
\parbox[t]{\textwidth} {
The decays $B_{d,s}^0 \rightarrow K^{(*)}
\bar{K}^{(*)}$ can be used to measure the angle $\alpha$ of the CKM
unitarity triangle. The theoretical error from $SU(3)$ breaking is
expected to be small, so that the determination of $\alpha$ is
clean. Moreover, since $B_{d,s}^0 \rightarrow K^{(*)} \bar{K}^{(*)}$
are pure penguin decays, they are particularly sensitive to the
presence of new physics.
}
%%
%\vskip2cm
\end{center}
\thispagestyle{empty}
\newpage
\setcounter{page}{1}
% Decrease texheight (for preprint numbers) again
%\textheight 23.0 true cm
\baselineskip=14pt
\section{CP phase $\alpha$ from $B_{d,s}^0 \rightarrow K^{(*)}
\bar{K}^{(*)}$}

The first evidence of CP violation in the $B$ system was recently
observed with the measurement of one of the angles of the
Cabibbo-Kobayashi-Maskawa (CKM) unitarity triangle: $\sin 2\beta =
0.78 \pm 0.08$ \cite{betameas}, which is consistent with the standard
model (SM). Future efforts will focus on the measurement of the
remaining two angles of the unitarity triangle, $\alpha$ and $\gamma$,
in order to test the SM explanation of CP violation.

There are two standard techniques for the extraction of $\alpha$. The
first method uses the CP asymmetry in $\bd(t)\to\pi^+\pi^-$ to obtain
$\alpha$. Unfortunately, there is a penguin contribution, making it
necessary to perform an isospin analysis of $B\to\pi\pi$ decays
\cite{isospin}. This requires the measurement of $\bd\to\pi^0\pi^0$,
which is expected to have a small branching ratio. Hence, it may be
difficult to obtain $\alpha$ using this method. The second method uses
a Dalitz-plot analysis of $\bd(t)\to\rho\pi\to\pi^+\pi^-\pi^0$ decays
\cite{Dalitz}. However, the unknown non-resonant background and the
correct description of $\rho\to\pi\pi$ decays are factors that can
seriously affect a clean determination of $\alpha$ using this method.

In this talk, we present a new method for determining $\alpha$
\cite{alpha}. As a starting point, consider the pure $b\to d$ penguin
decay $\bd \to K^0 \kbar$, for which the underlying quark transition
is ${\bar b} \to {\bar d} s {\bar s}$. The amplitude for $\bd \to K^0
\kbar$, $A_d$, can be written as
\bea
A_d & = & P_u V_{u}^d + P_c V_{c}^d +
P_t V_{t}^d  ~\nonumber\\
&=& (P_u-P_c)V_u^d+(P_t-P_c)V_t^d ~,
\label{BKKamp}
\eea
where $V_q^d \equiv V_{qb}^*V_{qd}$, and $P_{u,c,t}$ are the penguin
amplitudes. In passing from the first line to the second, we have
used the unitarity of the CKM matrix, $V_{ub}^* V_{ud} + V_{cb}^*
V_{cd} + V_{tb}^* V_{td} = 0$, to eliminate the $V_{cb}^* V_{cd}$
term. The amplitude ${\bar A}_d$ describing the conjugate decay
$\bdbar \to K^0 \kbar$ can be obtained from the above by changing the
signs of the weak phases.

By making time-dependent measurements of $\bd(t)\to K^0\kbar$, one can
obtain the three observables
\bea
X &\equiv & \frac{1}{2} \left( |A_d|^2 + |\bar{A}_d|^2 \right)
\nonumber\\
Y &\equiv & \frac{1}{2} \left( |A_d|^2 - |\bar{A}_d|^2 \right)\nonumber\\
\label{BKKobservables}
\ZI &\equiv & {\rm Im}\left( e^{-2i \beta} A_d^* {\bar A}_d \right).
\eea
The three independent observables depend on four theoretical
parameters: $P_{uc} \equiv |P_u-P_c|$, $P_{tc} \equiv |P_t-P_c|$, the
relative weak phase between the two amplitudes, $\alpha$, and the
relative strong phase. Hence one cannot obtain CP phase information
from these measurements \cite{LSS}. However, substituting
Eq.~\ref{BKKamp} in Eq.~\ref{BKKobservables}, one can obtain
\beq
P_{tc}^2|V_t^d|^2 = {\ZR \cos 2\alpha + \ZI \sin 2\alpha - X \over \cos
  2\alpha - 1} ~,
\label{singleeqn}
\eeq
where
\bers
\ZR & \equiv & {\rm Re}\left( e^{-2i \beta} A_d^* \bar{A}_d \right)~. \
\label{ZR} 
\eers
The quantity $\ZR$ is related to the three observables in
Eq. \ref{BKKobservables} by
\beq
\ZR^2 = X^2 - Y^2 - \ZI^2 ~.
\eeq

Now consider a second pure $b\to d$ penguin decay of the form $\bd \to
K^* \kstarbar$. Here $K^*$ represents the ground state vector meson,
$K^*(892)$, or any excited neutral kaon, such as $K_1(1270)$,
etc. This second decay can be treated in a similar fashion as the
first one above, with unprimed parameters and observables being
replaced by primed ones. One can then combine measurements of the two
decays to obtain
\bea
r_d \equiv {P_{tc}^2 \over P_{tc'}^2} & = & {\ZI \sin 2\alpha + \ZR
  \cos 2\alpha - X \over \ZI' \sin 2\alpha + \ZR' \cos 2\alpha -
  X'}\nonumber\\
&=& f(\alpha) ~.
\label{alphasolve}
\eea
The equation above, $r_d=f(\alpha)$, could then be solved for $\alpha$
if we knew $r_d$. Note that the CKM elements on the left-hand side of
Eq.~\ref{singleeqn} cancel in constructing the ratio $r_d$.

Information about the ratio $r_d$ can be obtained by measuring $\bs$
decays to the same final states $K^0\kbar$ and $K^* \kstarbar$.
Consider first the decay $\bs\to K^0\kbar$. This is described by a
$b\to s$ penguin amplitude, $A_s$, which is given by
\bea
A_s & = & P_u^{(s)} V_{u}^s + P_c^{(s)}
V_{c}^s + P_t^{(s)} V_{t}^s \nonumber\\
& \simeq & (P_t^{(s)} - P_c^{(s)}) V_{t}^s \ ~,
\label{Bsdecay}
\eea
where $V_q^s \equiv V_{qb}^*V_{qs}$, and $P_{u,c,t}^s$ are the penguin
amplitudes. In writing the second line, we have again used the
unitarity of the CKM matrix to eliminate the $V_{cb}^* V_{cs}$
piece. Furthermore, the $V_{ub}^* V_{us}$ piece is negligible:
$|V_{ub}^* V_{us}| \ll |V_{tb}^* V_{ts}|$. Thus, the measurement of
the branching ratio for $\bs\to K^0\kbar$ yields
$|P_t^s-P_c^s||V_t^s|$. Similarly one can obtain
$|P_t^{'s}-P_c^{'s}||V_t^s|$ from the branching ratio for $\bs \to K^*
\kstarbar$. In this way, we can measure 
\beq
r_s \equiv {{P_{tc}^{(s)}}^2 \over {P_{tc}^{'(s)}}^2} ~,
\eeq
where $P_{tc}^s \equiv |P_t^s-P_c^s|$ and $P_{tc}^{'s} \equiv
|P_t^{'s}-P_c^{'}s|$.

Now, the main point is that, up to small $SU(3)$ corrections,
\beq
r_d = r_s ~.
\label{Pratios}
\eeq
Note again that the CKM matrix elements cancel in both ratios. The
$SU(3)$ breaking in the equality $r_d=r_s$ is at the level of 5\% or
less \cite{alpha}. Hence $\alpha $ can be obtained from the equation $
r_d \approx r_s=f(\alpha)$. This equation has several solutions and so
there are discrete ambiguities in the extraction of $\alpha$.
However, by comparing several pairs of processes, the discrete
ambiguities can be eliminated. In fact, with one theoretical
assumption, all the discrete ambiguities can be removed with a single
pair of processes \cite{alpha}.

This method can also be used when the final state is not
self-conjugate. For example, one can consider the decays $\bd \to K^0
\kstarbar$ and $\bd \to K^{0*}\kbar$ \cite{alpha}.

{}From the above analysis, we therefore see that the CP phase $\alpha$
can be cleanly extracted from measurements of the decays of $\bd$ and
$\bs$ mesons to two different final states consisting of one neutral
kaon (i.e.\ $K^0$ or any of its excited states) and one neutral
anti-kaon (i.e.\ $\kbar$ or any excited state). Finally, we note that
the $K^* {\bar K}^*$ final state consists of three helicity
states. Each helicity state can be then considered a distinct final
state for the purposes of our analysis. Thus, by applying our method
to two different $K^* {\bar K}^*$ helicity states, $\alpha$ can be
obtained from $B^0_{d,s} \to K^* {\bar K}^*$ decays alone.

The branching ratios of the pure pure $b\to d$ penguin decays $\bd(t)
\to K^{(*)} {\bar K}^{(*)}$ are expected to be quite small, of order
$10^{-6}$. Hence this method is ideally suited to hadron colliders as
they produce an enormous number of $B$ mesons. Furthermore, in all
cases, the kaon or anti-kaon can be detected using its decays to
charged $\pi$'s or $K$'s only; this method does not require the
detection of $\pi^0$'s. Therefore hadron colliders will be able to use
this technique to measure $\alpha$ -- all that is required is good
$\pi$/$K$ separation. And if $\pi^0$'s can be detected, this simply
increases the detection efficiency for the various final states.

\section{New Physics in $B_{d,s}^0 \rightarrow K^{(*)} \bar{K}^{(*)}$  }

The decays $B_{d,s}^0 \rightarrow K^{(*)} \bar{K}^{(*)}$ are pure
penguin decays and hence could be sensitive to new-physics effects.
Consider the decays $B_{s} \rightarrow K^{(*)} \bar{K}^{(*)}$, which
are dominated to a very good approximation by a single amplitude in
the standard model (see Eq. \ref{Bsdecay}). Hence a measurement of CP
violation, such as a direct CP asymmetry, will be a clear sign of new
physics in the $b \rightarrow s $ penguin. 

New physics in the $ b \rightarrow s$ penguin can also affect the more
well known decay $B_d \rightarrow \phi K_S$ \cite{phiK}. However, note
that the new-physics operator for $B_{s} \rightarrow K^{(*)}
\bar{K}^{(*)}$ is of the form $O_d={\bar{d}}\Gamma_1 d \bar{s}
\Gamma_2 b$, where $\Gamma_{1,2}$ are some Lorentz operators, while
for $B_d \rightarrow \phi K_S$ the new-physics operator is of the form
$O_s={\bar{s}}\Gamma_1 s \bar{s} \Gamma_2 b$. There are models of new
physics where the operators $O_s$ and $O_d$ are related. For example,
consider models with an additional vector-singlet charge $-1/3$ quark
$h$ which mixes with the ordinary down-type quarks $d$, $s$ and
$b$. (These models are generally motivated by $E_6$ grand unified
theories.) This then generates flavour-changing effects through the
$Zb{\bar s}$ FCNC coupling \cite{NirSilv}. This coupling will then
generate the operators $O_{s,d}$ but with the same strength.
 
Models of new physics which contain exotic fermions generally have
additional neutral $Z'$ gauge bosons. If the $s$-, $b$- and $h$-quarks
have different quantum numbers under the new $U(1)$ symmetry, their
mixing will also induce FCNC's due to $Z'$ exchange
\cite{Z'FCNC,LerLon} which will then generate $O_{s,d}$, but again
with the same strength. Hence in such models CP violation in $B_d
\rightarrow \phi K_S$ and $B_{s} \rightarrow K^{(*)} \bar{K}^{(*)}$
will be correlated.

On the other hand, consider another model of new physics: R-parity
breaking supersymmetry (SUSY). The most general superpotential of the
MSSM with $SU(3)\times SU(2)\times U(1)$ gauge symmetry which breaks
R-parity is
\bea
{\cal W}_{\not \! R}&=&\frac{1}{2}\lambda_{[ij]k}L_iL_jE_k^c
+\lambda_{ijk}^{\prime}L_iQ_jD_k^c \nonumber\\
            & + & \frac{1}{2}\lambda_{i[jk]}^{\prime\prime}
U_i^cD_j^cD_k^c+\mu_iL_iH_2 ~.
\label{lag}
\eea
Here $L_i(Q_i)$ and $E_i(U_i,D_i)$ are the left-handed lepton (quark)
doublet and lepton (quark) singlet chiral superfields, where $i,j,k$
are generation indices and $c$ denotes a charge-conjugate field.
$H_{1,2}$ are the chiral superfields representing the two Higgs
doublets.  The non-observation of proton decay imposes very stringent
conditions on the simultaneous presence of both the baryon-number and
lepton-number violating terms in the Lagrangian \cite{Proton}.

The B-violating couplings $\lambda^{\prime \prime}$ are antisymmetric
in the last two indices. Hence the operator $O_s$ cannot be generated
at tree level and so there will no significant effects in $B_d
\rightarrow \phi K_S$. On the other hand, the operator $O_d$ can be
generated at tree level and hence can lead to CP violation in $B_{s}
\rightarrow K^{(*)} \bar{K}^{(*)}$.

The $L$-violating couplings are given in terms of four-component Dirac
spinors by \cite{DatXin}
\bea
{\cal L}_{\lambda^{\prime}}&=&\lambda^{\prime}_{ijk}
\left [\tilde e^i_L\bar d^k_R u^j_L
       +\tilde u^j_L\bar d^k_R e^i_L
       + {\tilde d^{k*}_R}{\bar e^{ic}_L} u^j_L
\right.\nonumber\\
& & \left.
-\tilde \nu^i_L\bar d^k_R d^j_L-\tilde d^j_L\bar d^k_R\nu^i_L
      -(\tilde d^k_R)^*(\bar \nu^i_L)^c d^j_L\right ]\nonumber\\
&+& h.c.\
\label{Lviolating}
\eea
In this case both operators $O_s$ and $O_d$ will be generated, but in
general with different strengths. Thus, in this model CP violation in
$B_d \rightarrow \phi K_S$ and $B_{s} \rightarrow K^{(*)}
\bar{K}^{(*)}$ can be quite different.

Finally, we return to the measurement of the CP phase $\alpha$ via
$B_{d,s}^0 \rightarrow K^{(*)} \bar{K}^{(*)}$. If the value of
$\alpha$ obtained via this method differs from that measured in $B_d
\to \pi \pi$ or $B_d \to \rho \pi$, this will be evidence of new
physics in the $b\to d$ or $b\to s$ penguin amplitudes.

\section{Conclusion}

In conclusion, we have a presented a new method to measure $\alpha$
using $B_{d,s}^0 \rightarrow K^{(*)} \bar{K}^{(*)}$. Because these
processes are pure penguin decays, they are particularly sensitive to
new physics. We have described several ways of detecting new physics
in such decays.

\bigskip
\noindent
{\bf Acknowledgements}:
%\bigskip
This work was financially supported by NSERC of Canada.

%%%%%%%%%%%%%%%%%%%%% REFERENCES %%%%%%%%%%%%%%%%%%%%%%%%%%%%%%%%

\end{document}